\documentclass{jfm}
\usepackage{amsmath,amsfonts,amssymb,bm,upmath,graphicx}

\title{Phase-field model for the Rayleigh--Taylor instability of 
immiscible  fluids}
\author[A. Celani, A. Mazzino, P. Muratore-Ginanneschi and L. Vozella]{
 A\ls N\ls T\ls O\ls N\ls I\ls O\ns  C\ls E\ls L\ls A\ls N\ls I$^1$\ns,
 A\ls N\ls D\ls R\ls E\ls A\ns M\ls A\ls Z\ls Z\ls I\ls N\ls O$^2$\ns, 
 P\ls A\ls O\ls L\ls O\ns M\ls U\ls R\ls A\ls T\ls O\ls R\ls E\ls-\ls G\ls I\ls N\ls A\ls N\ls N\ls E\ls S\ls C\ls H\ls I$^3$\ns
 and L\ls A\ls R\ls A\ns V\ls O\ls Z\ls E\ls L\ls L\ls A$^{2,3}$}
\affiliation{
 $^1$Institut Pasteur, CNRS, URA 2171, 25 Rue du docteur Roux, 75015 Paris, France\\[\affilskip]
 $^2$Department~of~Physics~-~University~of~Genova, and CNISM \& INFN~-~Genova~Section, via~Dodecaneso~33, 16146~Genova, Italy\\[\affilskip]
$^3$Department~of~Mathematics~and~Statistics~-~University~of~Helsinki, P.O. Box 4, 00014 Helsinki, Finland}
\date{\today}

\usepackage{natbib}
\allowdisplaybreaks[1]
				   
\begin{document}

\maketitle

\begin{abstract}
The Rayleigh--Taylor instability of two immiscible fluids in the limit of small
Atwood numbers is studied by means of a phase-field description.
In this method the sharp fluid interface is replaced by a
thin, yet finite, transition
layer where the interfacial forces vary smoothly. This is achieved by
introducing an order parameter (the phase field) whose variation is continuous
across the interfacial layers and is uniform in the bulk
region. The phase field model obeys a Cahn--Hilliard equation
and is two-way coupled to the standard Navier--Stokes equations.
Starting from this
system of equations we have first performed a linear analysis from which we
have analytically rederived the known gravity-capillary dispersion relation
in the limit of vanishing
mixing energy density and capillary width.
We have performed numerical simulations and identified a region of parameters
in which the known properties of the linear phase 
(both stable and unstable)
are reproduced in a very accurate way. This has been done both in the case
of negligible viscosity and in the case of nonzero viscosity. In the latter
situation only upper and lower bounds for the 
perturbation growth-rate are known.
Finally, we have also investigated the weakly-nonlinear stage of 
the perturbation evolution and identified a regime characterized by 
a constant terminal velocity
of bubbles/spikes. The measured value of the terminal velocity is in
perfect agreement with available theoretical prediction.
The phase-field approach thus appears to be a valuable tecnhique 
for the dynamical
description of the stages where hydrodynamic turbulence and 
wave-turbulence enter into play. 

\end{abstract}

\section{Introduction}
The Rayleigh-Taylor (RT) instability is a fluid-mixing
mechanism occurring when a heavy, denser, fluid 
is pushed into a lighter one. For a fluid in a gravitational field,
such a mechanism was first discovered by Lord Rayleigh in the 1880s 
\cite[][]{R883} 
and later applied to all accelerated fluids by
Sir Geoffrey Taylor in 1950 \cite[][]{T50}.
The relevance of this mixing mechanism embraces many different
phenomena occurring in completely different contexts. 
We just mention, among the many,
astrophysical supernova explosions and geophysical formations
like salt domes and volcanic islands \cite[][]{PS81,DS00}, 
continental magmatism caused by lithospheric gravitational instability
\cite[][]{LRB01,DS98}
, inertial confinement fusion \cite[][]{CZ02} 
and cloud formation in atmospheric sciences \cite[][]{Schultz}.\\
Back to classical fluids applications, RT instability is the first step
eventually leading to a fully developed turbulent regime.
A deeper understanding of the mechanism of flows driven by RT instability 
thus would shed light on the many processes that underpin fully developed 
turbulence. 
 
The difficulty inherent in sustaining an unstable density stratification 
has challenged experimentalists for over half a century. Several innovative 
approaches have been recently developed \cite[see e.g., ][]{RA04}.\\
With the advent of supercomputers, high-resolution numerical 
simulations of RT at high Reynolds numbers have become a reality. However, 
simulations using many different benchmark codes and 
experiments disagree 
already on apparently innocent observables like, 
for instance, the value of the 
growth constant, $\alpha$, associated to the spread of the 
turbulent mixing zone  \cite[see, e.g., ][]{PS81}. 
The differences can be as high as $100 \%$. 

Despite the long history of RT turbulence, a consistent phenomenological 
theory has been presented only very recently by \cite{C03} 
for the 
miscible case.  The theoretical predictions by Chertkov have been 
verified by \cite{CMV06} exploiting numerical simulations 
in two spatial dimensions. For the three-dimensional miscible 
case we refer, e.g, to \cite{RT3D}.

In many of the aforementioned situations where the RT instability
has an important role, the two fluids are immiscible owing to a non 
negligible surface tension.  At level of linear analysis the role played
by a non zero surface tension was addressed by \cite{C61}. The successive
 dynamics falling in a 
turbulent regime has been recently analyzed by \cite{C05}.
Using a phenomenological approach, the authors suggest
the existence of a  Kolmogorov cascade between the integral scale 
and a time-dependent scale related to the  
typical drop size. Below the latter scale, associated 
to an emulsion-like region, 
a wave energy cascade takes in. This is mediated by weakly 
interacting capillary waves propagating on top of the drop surface.
Eventually, the energy is dissipated by viscous forces.\\
RT instability and RT turbulence of immiscible fluids 
thus appear richer than the corresponding miscible situations.
The existence of two different cascades poses a serious challenge to 
numerical investigations of the immiscible RT problems. The emulsion-like
phase indeed occurs at very small scales and the energy transfer 
takes place on the interfaces. These are geometrical objects 
close to singularities and thus difficult to describe appropriately in
a numerical scheme.
Accuracy and efficiency are thus fundamental requirements 
to reproduce the correct statistical features characterizing 
immiscible RT turbulence. 

Our aim here is to perform a first step along this direction by
focusing on direct numerical simulations of immiscible RT instability.
The numerical strategy we exploit here is known 
as phase-field model \cite[][]{B02, CH58, BCB03, DSS07}.  
The main idea of the method is to treat
the interface between two immiscible fluids as a thin mixing layer 
across which physical properties vary steeply but continuously. 
The evolution of the mixing layer is ruled by an order parameter (the phase field)  that obeys a Cahn-Hilliard equation \cite[][]{CH58}. 
The method permits to 
avoid a direct tracking of the interface and easily 
produces the correct interfacial tension from the mixing layer free energy.

We present here an accurate numerical study that validates the phase--field approach by testing known results
of immiscible RT instability 
both at level of linear and weakly nonlinear analysis.
From our results, it turns out that this strategy is a valuable option
for a quantitative treatment of the turbulent regime characterized
by the interplay between hydrodynamic and interface degrees of freedom.

The paper is organized as follows.
In Sec.~\ref{pfm} we introduce the Rayleigh--Taylor problem and discuss
the related phase-field approach.
A detailed analysis
of the 
energy balance between purely hydrodynamic degrees of freedom and interface 
degrees of freedom is presented. Finally, the dispersion relation
for gravity-capillary waves is obtained by analytical calculations  
starting from  the phase-field equation coupled to the Navier--Stokes equations.
\\
In Sec.~\ref{num} the results from the direct numerical simulations are presented and 
compared with known results for the linear analysis. We focus both on the case of zero viscosity and on that of negligible viscosity. Both
stable and unstable configurations are considered.  
Finally, the weakly nonlinear
regime is considered and the resulting terminal velocity of bubbles/spikes
compared with existing theoretical predictions.\\
 Sec.~\ref{conclu} is devoted to some conclusions and perspectives.

\section{System configuration and phase-field model}\label{pfm}
\begin{figure}
\begin{center}
\includegraphics[scale=0.4]{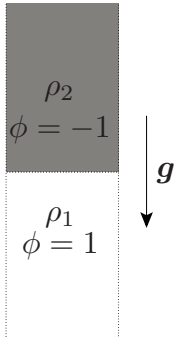}
\caption{\label{fig:fluids}Fluids configuration corresponding to a heavier fluid of density $\rho_2$ 
placed above a lighter one of density $\rho_1 < \rho_2$.}
\end{center}
\end{figure}

Our system  consists of two immiscible, incompressible fluids 
(labeled by 1 and 2) having different densities, 
$\rho_1$ and $\rho_2 > \rho_1$, 
with the denser fluid placed, e.g., above 
the less dense one (see Fig.~\ref{fig:fluids}). 
In the absence of gravity, 
this flow  configuration is stable. 
In presence of the gravitational force, surface tension may be able 
to keep the system in equilibrium, provided the density contrast 
is not too large.\\
Let us start by describe the equilibrium configuration and then 
pass  to the
evolution (RT instability) that occurs 
when a perturbation is imposed to the interface separating the two 
fluids. 

\subsection{Equilibrium state}
Let us consider an equilibrium state where fluid 1  is 
placed  below fluid 2 and they are separated  by a sharp 
interface. The fact that the interface is sharp (i.e. a discontinuity 
in the fluid properties) poses a serious challenge to numerical simulations. 
Indeed,
for sharp interfaces, the evolution equations are obtained 
by following fluid 1 and  2 separately with the appropriate boundary 
condition at interface \cite[see, for instance,][]{SHL05,S99}. 
Other approaches follow 
the interface alone. 
In this latter case, 
the movement of the interface is naturally amenable to 
a Lagrangian description, while the bulk flow is conventionally 
solved in an Eulerian framework. These approaches employ 
a mesh that has grid points on the interfaces 
and deforms according to the flow. A major shortcoming of these 
approaches is in that they cannot 
handle properly topological changes such as 
breakup, coalescence and reconnections 
\cite[see][and references therein]{YFLS04}.
In this respect, the phase--field  method is, by far and large, 
more effective, at the expense of a larger number of grid points required.

The idea of the phase--field method is to replace the sharp interface 
with a diffuse one in such a way 
that the numerical computation of interface movement and deformation 
can be carried out on fixed grids \cite[][]{AMW98, J99}. 
More 
quantitatively, this amounts to assigning to the system a Ginzburg--Landau 
free energy, $\mathcal{F}$, espressed in term 
of the order parameter $\phi$ as \cite[][]{CH58,B02,YFLS04}:
\begin{equation}
\mathcal{F}[\phi]=\int_{\Omega} \frac{\Lambda}{2}|\bm{\partial}\phi(\bm{x})|^2 + \frac{\Lambda}{4 \epsilon^2}(\phi^2 -1)^2
 d\bm{x}\quad,
\label{eq:freeenergy}
\end{equation}
where $\Omega$ is the region of space occupied by the system, 
$\Lambda$ is a mixing energy density
and $\epsilon$ is the capillary 
width, representative of the interface thickness. 
The order parameter $\phi$ is a field which 
serves to identify fluid 1 and 2. We assume 
$\phi=1$ in the region occupied by fluid 1 and $\phi=-1$ 
in those where fluid 2 is present.\\ 
The equilibrium state is the minimizer of the free energy $\mathcal{F}$. 
The mechanism which keeps the system in this configuration  is 
the competition between two effects due to the two addends
in \eqref{eq:freeenergy}. 
The first term
favours a perfect mixing (i.e. $\Lambda|\bm{\partial} \phi|^2/2 = 0$ in $\mathcal{F}$, 
this term being  the interface energy contribution) whereas the second one 
one drives the system towards demixing (the associated 
term in $\mathcal{F}$, the 
bulk contribution, has indeed a minimum for $\phi=\pm 1$).
The nontrivial final equilibrium 
state is just  the results of this competition.
More quantitatively, the final state is obtained by minimizing the free-energy 
functional with respect to variations of the function $\phi$, i.e., solving:  
\begin{equation}\label{eqequi}
\mu\equiv \delta\mathcal{F}/\delta\mathcal{\phi}=0\,\Leftrightarrow\,
-\partial^2\phi + \frac{\phi^3\,-\,\phi}{\epsilon^2}\,=\,0\quad, 
\end{equation}
where $\mu$ is the so-called chemical potential \cite[see, for instance,][]{CH58,B02,YFLS04}. 
If one considers
an one-dimensional 
interface, varying along the gravitational direction $y$, one easily finds 
the solution of Eq.~(\ref{eqequi}) as 
\cite[][]{CH58,B02,YFLS04}: 
\begin{equation}\label{phipr}
\phi(y)=\pm\tanh{\left(\frac{y}{\sqrt{2}\epsilon}\right)}\quad.
\end{equation}
This solution exists and is stable in all dimensions
although the decay rate of perturbations depends upon the 
 dimensionality \cite[][]{KKT05}. 
From (\ref{eqequi})   
one immediately realizes that the sharp-interface limit is 
obtained for $\epsilon\rightarrow 0$: 
in this case $\tanh{\left(y/(\sqrt{2}\epsilon)\right)}\rightarrow {\mathrm{sign}}(y)$.
Moreover, the surface tension $\sigma$ is equal 
to the integral of the free-energy density along the interface \cite[see, for example,][]{LL00}. 
For a plane interface, this integral yields \cite[][]{CH58,B02,YFLS04}: 
\begin{equation}\label{eqsigma}
\sigma=\frac{2\sqrt{2}}{3}\frac{\Lambda}{\epsilon}\quad.
\end{equation}
It is now easy to verify how the sharp interface limit is obtained:
it suffices to take the limits $\Lambda$ and $\epsilon$ to zero 
keeping $\sigma$ fixed to the value prescribed by surface tension  
 \cite[][]{LS03}. 

\subsection{Perturbation evolution}\label{subsec:perte}
Let us now suppose to impose a small perturbation on the (finite thickness) 
interface separating the two 
fluids. 
Such perturbation will displace the phase field from 
the previous  equilibrium configuration, which minimized the free-energy 
$\mathcal{F}$, to a new configuration for which in general,  
$\mu \neq 0$. 
The system will react so as to try to reach again an equilibrium 
configuration. In formulae:    
\begin{equation}
\frac{\partial \phi}{\partial t} + \bm{v}\cdot\bm{\partial}\phi=
\gamma\bm{\partial}^2 \mu= 
\gamma\,\Lambda\bm{\partial}^2\left[-\partial^2\phi \,+\,\frac{\left(\phi^3 - \phi \right)}{\epsilon^2}\right]\quad, \label{eq:phi}
\end{equation}
$\gamma$ being  the so-called mobility 
\cite[see, for instance,][]{B02,YFLS04}. 
Notice the presence of the Laplacian  
operator in front of $\mu$. 
Notice that the mass of each fluid
is conserved, as imposed by the physics of 
the problem under consideration.\\ 
The dynamics of the velocity field
is governed by the usual Boussinesq Navier-Stokes equations 
\cite[][]{KC01} plus 
an additional stress contribution arising at the interface where 
the effect of surface tension enters into play \cite[][]{B02,YFLS04,BBCV05}. 
The equations of motion are:    
\begin{eqnarray}
\left(\partial_t v_\alpha + \bm{v}\cdot\bm{\partial}v_\alpha\right)&=& 
-\frac{{\partial}_\alpha p}{\rho_o} + \nu\partial^2{v}_\alpha 
-\frac{\phi}{\rho_o}{\partial}_\alpha\frac{\delta\mathcal{F}}{\delta\phi}
+\frac{\rho'}{\rho_o}\,g_\alpha\label{eqv1}\\ 
\bm{\partial}\cdot\bm{v}&=&0\quad. \label{eqv2}
\end{eqnarray}
In the first equation $\rho_o=(\rho_1 + \rho_2)/2$ and $\nu$ is the kinematic viscosity. 
The quantity 
 $-\phi\bm{\partial}(\delta \mathcal{F}/\delta \phi)/\rho_o$ is the 
coupling term that accounts for capillary forces.
It is easy to verify that it  can be rewritten as 
$-\Lambda\left(\partial^2\phi\bm{\partial}\phi\right)/\rho_o$ 
plus a gradient term which can be absorbed into the pressure term. 
Finally, $\rho' g_\alpha/ \rho_o$ is the buoyancy 
contribution, $\rho'$ being the deviation of 
the actual density, $\rho$, 
from the mean density $\rho_o$: 
\begin{equation}\nonumber
\rho'=\rho - \rho_o \quad.
\end{equation}  
The buoyancy contribution 
can be rewritten in terms of $\rho_1$, $\rho_2$ and $\phi$ as: 
\begin{eqnarray}\label{eqbuoy}
\frac{\rho'}{\rho_o} g_\alpha&=&\frac{\rho - \rho_o}{\rho_o} g_\alpha =\nonumber\\ 
                 &=& \frac{\rho_1 \left(\frac{1+\phi}{2}\right) + 
\rho_2 \left(\frac{1-\phi}{2}\right) - \rho_o}{\rho_o}\,g_\alpha\nonumber\\
                 &=& -\,\mathcal{A}\phi\,g_\alpha
\end{eqnarray} 
where $\mathcal{A}\equiv(\rho_2-\rho_1)/(\rho_2+\rho_1)$ is the Atwood number.

\subsection{Energetics}
Let us define the kinetic energy (per unit volume), $E_K$, and the potential
energy (per unit volume), $E_P$, for our system ruled by Eqs.~
(\ref{eq:phi}), (\ref{eqv1}) and  (\ref{eqv2}).\\ 
By definition of potential energy, we have: 
\begin{eqnarray}\label{eq:E_p}
E_P &=& \frac{1}{\Omega}\int\int dx\,dy\,\rho_2\,g\,y \frac{1-\phi}{2} + 
\frac{1}{\Omega}\int\int dx\,dy\,\rho_1\,g\,y \frac{1+\phi}{2} + E_P^o=\nonumber\\
   &=& -\frac{1}{2}\langle y\,\phi \rangle (\rho_2-\rho_1) g = - \rho_o\,\mathcal{A}\,g\,
\langle y\,\phi \rangle\quad,
\end{eqnarray}
 $\Omega$ being the total volume occupied  by the fluids and brackets, 
$\langle\cdots\rangle$, denote spatial averages. In Eq.~({\ref{eq:E_p}}) 
the constant $E_P^o$ is chosen such to set the potential energy to zero for
vanishing Atwood number.\\
In a similar way, one can define the kinetic energy per unit volume as:
\begin{eqnarray}
E_K &=& \frac{1}{\Omega}\int\int dx\,dy\,\rho_2\,\frac{1-\phi}{2}\,\frac{\bm{v}^2}{2} + 
\frac{1}{\Omega}\int\int dx\,dy\,\rho_1\,\frac{1+\phi}{2}\,\frac{\bm{v}^2}{2} \,=\nonumber\\
   &=& {\rho_2}\langle\left(\frac{1-\phi}{2}\right)\,\frac{\bm{v}^2}{2}\rangle\,+\,
{\rho_1}\langle\left(\frac{1+\phi}{2}\right)\,\frac{\bm{v}^2}{2}\rangle  \,= \nonumber\\
   &=&\rho_o\langle\frac{\bm{v}^2}{2}\rangle \,-\, \rho_o\,\mathcal{A}\langle\phi\,\frac{\bm{v}^2}{2}\rangle\quad.
\end{eqnarray}
From Eqs.~(\ref{eq:phi}) and (\ref{eqv1}) we immediately 
realize that such equations are left invariant under the simultaneous 
transformation $\bm{g}\rightarrow -\bm{g}$, $\phi\rightarrow-\phi$. 
As a consequence, $\langle\phi\,\bm{v}^2/2\rangle \,=\!0$ and the resulting kinetic 
energy simply reads:
\begin{equation}
E_K = \rho_o\,\langle\frac{\bm{v}^2}{2}\rangle\quad.
\end{equation} 
By defining $E_\mathcal{F}\equiv\mathcal{F}/\Omega$ 
the total energy of the two-fluid system is 
$$
E = E_P +E_K +E_{\mathcal F}\quad.
$$
The equation for $E_K$ is obtained by 
multiplying Eq.~(\ref{eqv1}) by $\rho_o v_\alpha$ and 
then taking spatial average. We easily get: 
\begin{equation}\label{eqek_t}
d E_K/dt =
\rho_o\partial_t \langle\frac{\bm{v}^2}{2}\rangle = 
- \rho_o \nu \langle\left(\partial_\alpha\,\bm{v} \right)^2 \rangle 
+ \rho_o \,\mathcal{A}\, g \langle v \phi\rangle - 
\Lambda \langle v_\alpha\left(\partial_\alpha \phi\right)\left(\partial^2 \phi\right) \rangle\quad.
\end{equation}
Let us now take Eq.~(\ref{eq:phi}), multiply it by 
$y$, and take the average: 
\begin{equation}
\partial_t \langle y\,\phi\rangle + \langle y\,\partial_y\left(v\phi\right)\rangle\,=\, 
\gamma\Lambda \langle y\,\partial^2\left(-\partial^2 \phi\,+\,\frac{\phi^3 -\phi}{\epsilon^2} \right)\rangle\,= 0\quad,\\
\end{equation}
by translational invariance and Leibniz rule.  
We thus have: 
\begin{equation}\label{eqphi_t}
d E_P/dt = -\partial_t\left(\rho_o\mathcal{A} g\langle y\,\phi \rangle \right)\,=-\,\rho_o\mathcal{A} g \langle v\phi \rangle \quad, 
\end{equation}
where we have used the fact that $\langle y\partial_y (v\phi)\rangle = -\langle (\partial_y y)v\phi\rangle = -\langle v\phi\rangle$.
The free--energy variation is 
\begin{eqnarray}
\partial_t\mathcal{F} &=& \int\int \frac{\delta \mathcal{F}}{\delta \phi} 
\frac{\partial \phi}{\partial t} dx\,dy =\nonumber\\
&=&\int\int\frac{\delta\mathcal{F}}{\delta \phi}\left[-\bm{v}\cdot\bm{\partial}\phi+\gamma\partial^2\left(\frac{\delta\mathcal{F}}{\delta\phi}\right)\right] dx\,dy=\nonumber\\
&=&-\gamma\langle\left[\partial\left(\frac{\delta \mathcal{F}}{\delta\phi}\right)\right]^2\rangle\Omega 
- \int\int \left(\frac{\delta\mathcal{F}}{\delta\phi}\right)v_i\partial_i\phi\, 
dx\,dy=\nonumber\\
&=&-\gamma\langle\left[\partial\left(\frac{\delta\mathcal{F}}{\delta\phi}\right)\right]^2\rangle\Omega - 
\Lambda\int\int\left[(-\partial^2\phi)+\frac{\phi^3-\phi}{\epsilon^2}\right]
v_i\partial_i\phi\,dx\,dy=\nonumber\\
&=&-\gamma\langle\left[\partial\left(\frac{\delta\mathcal{F}}{\delta\phi}\right)\right]^2\rangle\Omega 
- \Lambda\langle e_{\alpha\beta}(\partial_\alpha\phi)(\partial_\beta\phi)\rangle\Omega\quad. \end{eqnarray}
i.e.,
\begin{equation}\label{eqfree_t}
\partial_t E_\mathcal{F} \,=\,-\gamma \langle\left[\partial_\alpha\left(\frac{\delta \mathcal{F}}{\delta\phi}\right)\right]^2\rangle \,-\,\Lambda\langle e_{\alpha\beta}\left(\partial_\alpha\phi\right)\left(\partial_\beta\phi\right)\rangle\quad. 
\end{equation} 
where we have introduced the strain tensor $e_{\alpha\beta}\equiv\left(
\partial_\alpha v_\beta + \partial_\beta v_\alpha \right) / 2$ and assumed boundary 
conditions suitable to justify integrations by parts.\\
The energy balance takes then the form: 
\begin{equation}
\partial_t(E_K\,+\,E_P\,+\,E_\mathcal{F})\,=\,-\rho_o\nu\langle\left(\partial_\alpha \bm{v}\right)^2\rangle
-\gamma\langle\left[\partial_\alpha\left(\frac{\delta\mathcal{F}}{\delta\phi}\right)\right]^2\rangle\quad.
\label{eq:etot}
\end{equation}
The global system in thus intimately dissipative, 
even for a vanishing kinetic viscosity.\\
It is worth emphasizing the cancellation of 
$\Lambda \langle e_{\alpha\beta}(\partial_\alpha\phi)(\partial_\beta\phi)\rangle$ 
by the kinetic and the free--energy contributions, due to 
exchanges between the velocity field and the interface. 
\subsection{Dispersion relation for the phase-field model}\label{subsec:dispe}
The aim of this section is to show that the well-known 
dispersion relation for gravity-capillary waves \cite[][]{C61} can 
be easily obtained within the phase-field formalism. To do that, 
let us 
concentrate our attention on a two-dimensional problem and indicate
by $y$ the gravity direction. 
Moreover, we will assume heavier fluid to be placed below the lighter 
one, in a way to have a stable situation.  
For a given  perturbation imposed to the interface, 
the problem is to determine how the perturbation evolves in time.\\
Denoting by $h(x,t)$ a small perturbation imposed to a planar interface, we
can rewrite $\phi$ as:
\begin{equation}
\phi = f\left(\frac{y-h(x,t)}{\epsilon}\right)\quad,
\end{equation}
where $h$ can be larger than $\epsilon$, yet it has to
be smaller than the scale 
of variation of $h$ (small amplitudes).\\  
Locally, the interface is in equilibrium, i.e.:
\begin{equation}
f'' = V'(f)\quad,
\end{equation}
where $V(\phi)=(\phi^2-1)^2/4 \epsilon^2$. 
In this limit we have:
\begin{equation}
\mu = - \Lambda \frac{\partial^2 f}{\partial x^2} = 
\frac{\Lambda}{\epsilon}\left [f' \frac{\partial^2 h}{\partial x^2} 
  - \frac{f''}{\epsilon} \left(\frac{\partial h}{\partial x}\right)^2\right ]\quad.
\end{equation}
Linearizing Eq.~(\ref{eqv1}) for small interface
velocity we have, neglecting the viscous term:
\begin{equation}
\rho_o \partial_t v = - \partial_y p - \phi \partial_y \mu - \mathcal{A} g \rho_o \phi\quad.
\end{equation}
The integration in the vertical direction interpreted 
in the principle value sense 
\begin{eqnarray}
q_y &:=& \lim_{L\uparrow\infty} \int_{-L}^{L} v \,dy\quad,\\
\rho_o \partial_t q_y &:=& \lim_{L\uparrow \infty}
\left\{
 \frac{\Lambda}{\epsilon} \int_{-L}^{L} \left[ 
f f''  \frac{\partial^2 h}{\partial x^2}
- \frac{1}{\epsilon} f f''' \left(\frac{\partial h}{\partial x}\right)^2
\right] d(y/\epsilon) - \mathcal{A} g \rho_o \int_{- L}^{L} f dy \right\} \quad,
\end{eqnarray}
yields:
\begin{equation}
\rho_o \partial_t q_y =
\sigma \frac{\partial^2 h}{\partial x^2} - 2 \mathcal{A} g \rho_o h\quad,
\end{equation}
having used the
relations $\int (f')^2 dy = 2 \sqrt{2}/3$, 
$\int f f''' dy = 0$ and  
\begin{equation}
\lim_{L\uparrow \infty} \int_{-L}^{+L} f dy = +2 h \quad. 
\end{equation}
The height variation of the interface has to match the vertical fluid velocity,
thus giving:
\begin{equation}
\partial_t h = v(x,h(x,t),t) \equiv v^{(int)}(x,t)\quad.
\end{equation}
The last step is to relate the velocity at the interface with the 
integral $q_y$. This is done by restricting to potential flows:
\begin{equation}
{\bm v} = {\bm \partial} \psi \qquad \partial^2 \psi =0\quad.
\end{equation}
For $y>0$, denoting with ``$\hat{\phantom{p}}$'' the Fourier Transform,  
we have:
\begin{equation}
\psi(x,y,t)=\int_0^{\infty} e^{-ky+ikx} \hat{\psi}(k,t) dk + \mathrm{c. c.}
\end{equation}
\begin{equation}
v(x,y,t) = -\int_0^{\infty} k e^{-ky+ikx} \hat{\psi}(k,t) dk + \mathrm{c. c.}
\end{equation}
\begin{equation}
q_y(x,t)= -2  \int_0^{\infty} e^{ikx} \hat{\psi}(k,t) dk + \mathrm{c. c.}
\end{equation}
\begin{equation}\label{eq:v_int}
v^{(int)} =  -\int_0^{\infty} k e^{ikx} \hat{\psi}(k,t) dk + \mathrm{c. c.}
\end{equation}
Therefore: 
\begin{equation}
\hat{v}^{(int)} =  \frac{k \hat{q}_y}{2}\quad,
\end{equation}
so that in $k-$space we have:
\begin{equation}
\partial_t \hat{h} = \frac{k \hat{q}_y}{2} \qquad 
\rho_o \partial_t \hat{q}_y = (- \sigma k^2 - 2 \mathcal{A} g \rho_o) \hat{h}\quad.
\end{equation}
From these two equations we immediately get:
\begin{equation}\label{eq:h}
\partial_t^2 \hat{h} + \omega^2 \hat{h} = 0\quad,
\end{equation} 
with:
\begin{equation}\label{eq:om2}
\omega^2(k)=+\mathcal{A}g k+\frac{\sigma}{2\rho_o}k^3
\end{equation}
that is the expected dispersion relation \cite[][]{C61}. 
 For the stable configuration we have, for all values of $\sigma$:   
\hbox{${\mathcal{A}g k+{\sigma}/\left({2\rho_o}\right)k^3} > 0$}, 
i.e. any initially imposed perturbation will not grow indefinitely.\\ 
From Eq.~(\ref{eq:h}) and the initial condition:  
\begin{equation}
\partial_t \hat{h}(k,t)=0 \quad \textrm{at}~t=0\quad,
\end{equation}
 we immediately have:
\begin{equation}\label{eq:h_t}
\hat{h}(k,t)=\hat{h}(k,0)\cos{(\omega t)}
\end{equation}
and the  velocity at the interface reads:
\begin{equation}\label{eq:v_int_2}
\hat{v}_y^{int}(k,t)=- \hat{h}(k,0)\omega\sin{(\omega t)}\quad. 
\end{equation}
Assuming an initial perturbation of the form $h(x,0)=h_0\cos{(\bar{k}x)}$, 
from Eqs.~(\ref{eq:v_int}) and (\ref{eq:v_int_2}) we obtain: 
\begin{equation}
\hat{\psi}(\bar{k},t)=\frac{1}{\bar{k}}\hat{h}(\bar{k},0)\omega\sin(\omega t)\quad,
\end{equation}
and the velocity components, for $y>0$,  read: 
\begin{eqnarray}
v^{\uparrow}(x,y,t)\equiv v(x,y,t)&=& - \cos{(\bar{k}x)}e^{-\bar{k}y}{h}_o\omega\sin{(\omega t)}\label{eq:vs_u}\\
u^{\uparrow}(x,y,t)\equiv u(x,y,t)&=& \phantom{+} \sin{(\bar{k}x)} e^{-\bar{k}y} {h}_o\omega\sin{(\omega t)}\quad,\label{eq:us_u}  
\end{eqnarray}
where  we used the relation $h_0=2\hat{h}(k,0)$.\\
For $y < 0$, in a similar way we obtain the velocity field components: 
\begin{eqnarray}
v^{\downarrow}(x,y,t)\equiv v(x,y,t)&=&-  \cos{(\bar{k}x)} e^{+\bar{k} y} {h}_o\omega\sin{(\omega t)}\label{eq:vs_d}\\
u^{\downarrow}(x,y,t)\equiv u(x,y,t)&=&-  \sin{(\bar{k}x)} e^{+\bar{k} y} {h}_o\omega\sin{(\omega t)}\quad.\label{eq:us_d}
\end{eqnarray}
When  
in the initial configuration the heavier fluid placed 
above the lighter one, the dispersion relation (\ref{eq:om2}) 
trasforms in: 
\begin{equation}\label{eq:inst}
\omega^2(\bar{k})=-\mathcal{A}g\bar{k}+\frac{\sigma}{2\rho_o}\bar{k}^3\quad, 
\end{equation} 
which is readly obtained by flipping the sign of $g$. 
For $\sigma < \sigma_c \equiv 2\rho_o / (\mathcal{A}g \bar{k}^2)$  
surface tension is not able to contrast gravity-induced vertical 
motion with the final result that amplitude perturbations 
grows exponentially: the flow is unstable. More precisely, from relation 
(\ref{eq:inst}) and for $\sigma < \sigma_c$ we have: 
\begin{equation}\label{eq:alpha}
\omega(\bar{k})=\sqrt{-\mathcal{A}g\bar{k} + \frac{\sigma}{2\rho_o}\bar{k}^3}\equiv i \alpha(\bar{k})\quad,
\end{equation}
and Eqs.~(\ref{eq:vs_u}) - (\ref{eq:us_d}) transform in:
\begin{eqnarray}
v^{\uparrow}(x,y,t)\equiv v(x,y,t)&=&  \phantom{+}\cos{(\bar{k}x)}e^{-\bar{k}y}{h}_0\alpha\sinh{(\alpha t)}\label{eq:vi_u}\\
u^{\uparrow}(x,y,t)\equiv u(x,y,t)&=& - \sin{(\bar{k}x)} e^{-\bar{k}y} {h}_0\alpha\sinh{(\alpha t)}\quad,\label{eq:ui_u}
\end{eqnarray}
for $y > 0$, and: 
\begin{eqnarray}
v^{\downarrow}(x,y,t)\equiv v(x,y,t)&=& \phantom{+} \cos{(\bar{k}x)} e^{+\bar{k} y} {h}_0\alpha\sinh{(\alpha t)}\label{eq:vi_d}\\
u^{\downarrow}(x,y,t)\equiv u(x,y,t)&=& \phantom{+} \sin{(\bar{k}x)} e^{+\bar{k} y} {h}_0\alpha\sinh{(\alpha t)}\quad,\label{eq:ui_d}
\end{eqnarray}
for $y < 0$.
\section{Numerical investigation}\label{num}
In this section we report results  we have  obtained exploiting 
direct numerical simulations (DNS) of the phase-field
model for the Rayleigh--Taylor problem
described in the preceeding sections. 
Our attention will be focused
both on the linear phase of the perturbation evolution and on the
weakly nonlinear regime governed by plumes, for $\mathcal{A} \ll 1$.  \\
In the present study we will
consider initial perturbations imposed to the interface varying along
one of the horizontal directions, say the $x$-axis, and invariant along the 
other horizontal direction, say the $z$-axis. 
The perturbation is thus intimately two-dimensional
a fact that allows us to solve the original Navier--Stokes equations coupled 
to the phase field in two dimensions. This clearly permits to obtain
high accuracy and thus to properly test the phase-field approach against
known results for both the linear and the nonlinear evolution stage.\\
For a two-dimensional flow it is 
convenient to introduce the vorticity field $\omega$ 
[$\omega = (\bm{\partial}\times\bm{v})_z$] 
and study 
the equations 
\begin{equation}
\partial_t \omega + \bm{v}\cdot \bm{\partial}\omega = 
+ \nu\partial^2 \omega -
\frac{\Lambda}{\rho_o}\bm{\partial}\times\left(\partial^2\phi\,\bm{\partial\phi}\right)
 - \mathcal{A}\left(\bm{\partial}\phi\right) \times \bm{g}
\label{eqomega}
\end{equation}
\begin{equation}
\partial_t \phi + \bm{v}\cdot\bm{\partial}\phi=
\gamma\bm{\partial}^2 \mu= 
\gamma\,\Lambda\bm{\partial}^2\left[-\partial^2\phi \,+\,\frac{\left(\phi^3 - \phi \right)}{\epsilon^2}\right]\quad. 
\label{eqphi}
\end{equation}
In order to efficiently and accurately solve those equations we exploit a 
pseudospectral method \cite{C88}. Accordingly, periodic boundary conditions
have to be assumed along the two directions. For the horizontal direction
it is a natural choice (see e.g. \cite{CC06,LS03}) 
while along the vertical one
this choice deserves some comments.  
As initial condition we started from the 
hyperbolic-tangent profile, Eq.~(\ref{phipr}), 
for $\phi$ with 
the interface placed in the middle of the 
domain. 
The fact that we have periodic boundary conditions along 
$y$ simply means that far from the middle of the domain the 
hyperbolic-tangent profile 
has to be distorted in order to satisfy periodic boundary conditions.
However, both in the linear and in the weakly nonlinear 
regimes the amplitude of the interface perturbation is always much 
smaller than 
the vertical size of the box, so that the actual choice of boundary conditions at the top and  bottom can be safely neglected. \\
Such a strategy  has been  
already exploited for the miscible case by \cite{CMV06}.\\
The box has a horizontal to vertical 
aspect ratio $L_x/L_y = 1$ for the  linear analysis stage and $L_x/L_y = 1/2$ 
for the weakly nonlinear evolution. 
In the latter case we take a smaller aspect ratio owing to the fact that
the perturbation can reach a higher amplitude (with respect to case of the 
linear analysis).\\
In both cases the resolution is $1024\times1024$ collocation 
points. 
We need such a high resolution (despite the fact that we focus on
a linear and weakly nonlinear study)
in order to have a well described interface separating the two phases. 
In our simulations the mixing width ($\sim 4\,\epsilon$) is 6 mesh points.\\ 
The time evolution is implemented by a standard second-order 
Runge--Kutta scheme.\\
The physically relevant parameters in the present problem
are the kinematic viscosity 
$\nu$, the buoyancy intensity $\mathcal{A}g$ and the surface tension $\sigma$.
Both $\mathcal{A}g$  and $\nu$ will be 
varied in our study, while $\sigma$ will be kept fixed to a fixed value (see
below).
The surface tension is related to the ratio $\Lambda/\epsilon$ with $\epsilon$ 
(and thus $\Lambda$)
sufficiently small in order to have a finite value for the
surface tension and, at the same time,
to reproduce the correct sharp-interface limit. 
Finally, the parameter $\gamma$  appearing in the relaxation term in 
Eq.~(\ref{eqphi}) must satisfy the requirement that $\gamma \Lambda$ be 
small, so as to enforce
`istantaneous' local equilibrium between flow and interface. Here we used
the value (model units)  $\gamma \Lambda = 10^{-8}$.

All simulations presented here 
start from an initial condition corresponding 
to an equilibrium configuration: velocity identically zero 
and  hyperbolic tangent profile for the phase field $\phi$, 
expressed by the relation of the form: 
$\tanh{((y - h(x,t=0))/ c)}$ with 
\begin{equation}\nonumber
h(x,\mathrm{t=0})=h_0\sin{(k\,x)}\quad.
\end{equation}
For a given  $k$ we choose the initial 
amplitude $h_0$ in a way that $h_0\,/ \,\lambda$ (where $\lambda \equiv 2\pi/k$) 
is sufficiently small to fall in the linear phase (i.e. $h_0\,/ \,\lambda \ll 1 $ )
and  $h_0$ is sufficiently large
for the wave disturbance to see an almost infinitesimal mixing width
(i.e. $h_0\,/ \,\epsilon \gg 1 $ ). Specific numerical values are reported in the 
next sections.


\subsection{Linear instability for negligible viscosity}
\label{num_ins}
\begin{figure}
 \centering 
 \includegraphics[scale=0.5]{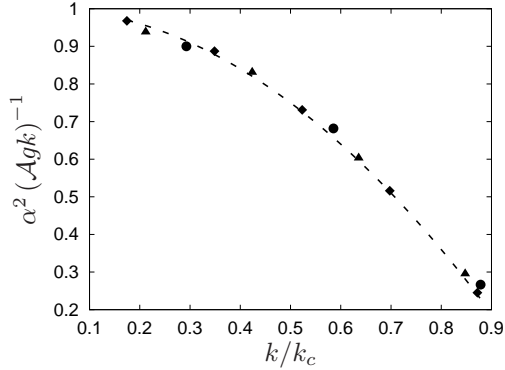}
\caption{\label{fig:alpha1}The square growth-rate $\alpha^2$ (see Eq.~(\ref{eq:alpha})) for three different values of $\mathcal{A}g$ corresponding 
to three different values of the critical wave number 
$k_c\equiv (2 \mathcal{A}g \rho_o / \sigma)^{1/2}$:  
$k_c=3.4$ (solid circle), $k_c=4.7$ (solid triangle) and 
$k_c=5.7$ (solid rhombus).  
The dashed line is the linear-theory prediction expressed by the relation 
(\ref{eq:alpha}).}
\end{figure}

\begin{figure}
\centering
\includegraphics[scale=0.5]{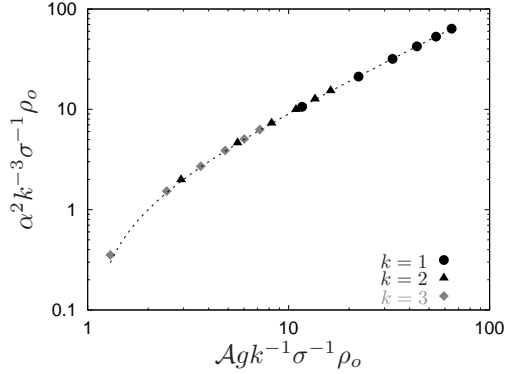}
\caption{\label{fig:alpha2} The square growth-rate $\alpha^2$ for $k=1$ (solid circles), $k=2$ (solid triangles) and 
$k=3$ (solid rhombus), all smaller than $k_c$,
for six different values of $\mathcal{A}g$ ranging from $0.11$ to $0.61$. 
The dashed line corresponds to the linear-theory  prediction.}
\end{figure}

\begin{figure}
\centering
\includegraphics[scale=0.6,angle=270]{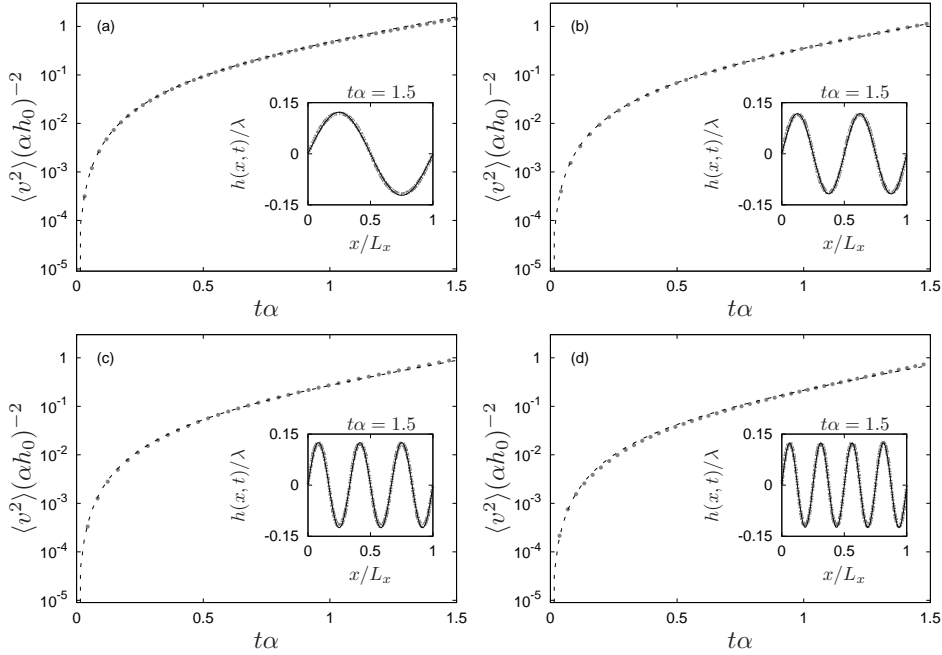}
\caption{\label{fig:w2}
Time behavior of $\langle{v^2}\rangle$  
for  $k_c = 4.7$ 
(in Fig.~\ref{fig:alpha1} corresponding to the solid triangle) and for four
values of $k<k_c$.
(a) $k=1$, (b) $k=2$, (c) $k=3$ and (d) $k=4$. The numerical results (symbols) are 
 compared with the corresponding best fit expressions 
(see the text for details). In the 
insets the interface perturbation, $h(x,t)$, is plotted at $t\alpha = 1.5$
revealing a very accurate linear analysis prediction.}
\end{figure}

The aim of this section is to verify the growth-rate (\ref{eq:alpha})
which holds in the linear phase when the viscosity is negligible. \\
In order to do so, we take a small value of $\nu$ ($\nu= 10^{-5}$ in the model units)
and vary $k$ (up to $k_c\equiv (2\mathcal{A}g \rho_o / \sigma)^{1/2}$,
the critical wave-number separating unstable from stable wave-modes) and  $Ag$
and take a fixed value of $\sigma$. The ratio $h_0\,/\,\lambda = 0.06$ while
$h_0\,/\,\epsilon$ ranges from $\sim 10$ to $\sim 40$ in the range of
$k$ considered.\\
The behavior of the square growth-rate $\alpha^2$  
is shown in dimensionless form in Fig.~\ref{fig:alpha1} 
as a function of $k$ for three different values 
of  $k_c$ (obtained by varying $Ag$)
and in Fig.~\ref{fig:alpha2} by varying $Ag$ for three different values of 
$k<k_c$. In both figures, symbols refer to the numerical results and 
the dashed line is the theoretical expectation given by (\ref{eq:alpha}). \\
The numerical data in Figs.~\ref{fig:alpha1} and \ref{fig:alpha2} have been obtained via best-fit 
of $\langle v^2\rangle$, the spatial average of $v^2$ as a function of time.  
The latter average is computed over 
a horizontal strip containing the interface (placed in the middle of the 
computational domain) and having an extension of $a_y$ above and below
the interface. This has been done to avoid spurious contaminations
coming from the upper and lower domain regions affected by the 
boundary conditions. In formulae:  
\begin{multline}\label{eq:w2}
\langle{v^2}\rangle = 
\frac{1}{2 a_y}\frac{1}{L_x}\int_{-a_y}^{0} dy \int_0^{L_x} dx \,{\left(v^{\downarrow}\right)^2}  
+ \frac{1}{2 a_y}\frac{1}{L_x} \int_{0}^{a_y} dy \int_0^{L_x} dx\,{\left(v^{\uparrow}\right)^2}\\
= \frac{1}{2\,a_y\,k}\left[ -e^{-2\,k\,a_y} + 1 \right]\alpha^2 h_0^2\sinh^2{(\alpha\,t)}\quad,
\end{multline}  
where we  used the expression (\ref{eq:vi_u}) and (\ref{eq:vi_d}) 
for $v^{\uparrow}$ and $v^{\downarrow}$, respectively.  \\
The best fit has been done with $\alpha $ as unique free parameter
and its high accuracy can be verified in  Fig.~\ref{fig:w2}
where we show the time evolution of $\langle{v^2}\rangle$ for $k_c=4.7$ 
(solid triangles in Fig.~\ref{fig:alpha1}) and for four values of $k$ smaller than $k_c$.
At  $t\alpha > 1.5$ nonlinear
effects 
start to enter into play giving rise to corrections to the linear analysis
(see Sec.~\ref{num_nl}). Up to that time, linear theory is very accurate as
one can also realize by looking at the insets of Fig.~\ref{fig:alpha1}
where the sinusoidal form of $h(x,t)$ is reported for  $t\alpha = 1.5$.

\subsection{Linear instability for finite viscosity}
\label{num_nu}
\begin{figure}
\centering
\includegraphics[scale=0.5]{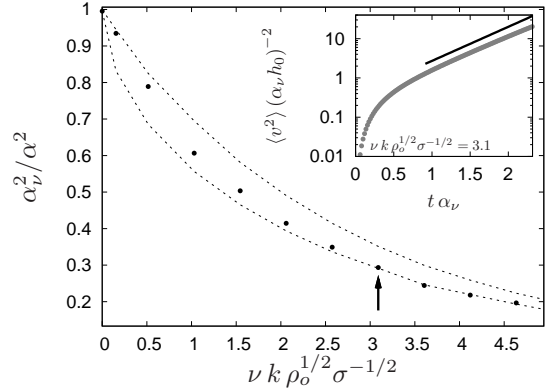}
\caption{\label{fig:nu}
Behavior of the dimensionless perturbation growth-rate, $\alpha_{\nu}$, for 
$k=1$  and $\mathcal{A}g$ corresponding to $k_c =5.7 $. 
Dotted lines correspond
to upper and lower bounds for the growth-rate (see Eqs.~(\ref{eq:up}) and (\ref{eq:low})). 
The arrow selects a value of the  viscosity for which the 
time evolution of $\langle v^2 \rangle$ is reported in the inset.
The  continuous  line is the best fit slope (see text).
}
\end{figure}

The aim of this section is to investigate numerically 
how the growth-rate, $\alpha$, is modified by
viscosity. As discussed in Appendix \ref{sec:l_vu}, both an upper and a lower bound
for the perturbation growth-rate are known (see Eqs.~(\ref{eq:low}) and (\ref{eq:up})) and we 
want to assess how the actual growth-rates compare with those.\\
For such purpose,
we choose a  surface tension, $\sigma$, and $\mathcal{A}g$
in such a way to obtain instability for
few (unstable) wavenumbers. Our choice was 
$k_c=5.7$ (see Sec.~\ref{num_ins})
thus corresponding to 5 unstable wavenumbers. \\
As far as the initial perturbation is concerned, we report here the case corresponding
to $k=1$.  Initial perturbations with a larger wavenumber simply
need an initial smaller amplitude (and eventually a larger numerical resolution)
in order to satisfy
 $h_0\gg\epsilon$  and  $h_0\ll \lambda$. Here, we have
$h_0 / \lambda = 0.03$ and $h_0 / \epsilon \sim 20$. Such ratios turned out to be sufficiently
`asymptotic' to produce accurate results.
 The effect of viscosity is studied
by considering twelve values of viscosity in the range $10^{-5}\le \nu \le 5~10^{-2}$ (model units).\\
The results of our simulations are summarized in Fig.~\ref{fig:nu} where the behavior of the square 
 perturbation growth-rate, $\alpha_{\nu}^2$, 
is shown as a function of viscosity. The numerical predictions have been compared with the 
available theoretical bounds (dashed lines).\\
Note that the numerical points are always in between the two bounds and also how the relative differences
between the upper bound and the numerical values are $< 11\%$. This latter fact is compatible, for example,
with the results of \cite{MMSZ77}.\\
The value of the growth-rates have been obtained via best of $\langle v^2 \rangle$ (see Eq.~(\ref{eq:w2})).
Unlike what we did in previous section, here we  perform the fit within the  exponential region. 
The reason is that the non-asymptotic form of the perturbation time-evolution is unknown in the present case.

The fit accuracy can be appreciated in the inset of Fig.~\ref{fig:nu}
where the temporal evolution of the pertubation for $\nu=0.3$ (model units) is shown together with the best fit slope
(dashed line) from which $\alpha_v$ is determined.
Error bars, estimated by looking at the fit sensitivity by varying the length of the fit interval, 
are of the order of the symbol sizes. \\

\subsection{Stable configuration: gravity-capillary waves}
\label{num_sta}
\begin{figure}
\centering
\includegraphics[scale=0.5]{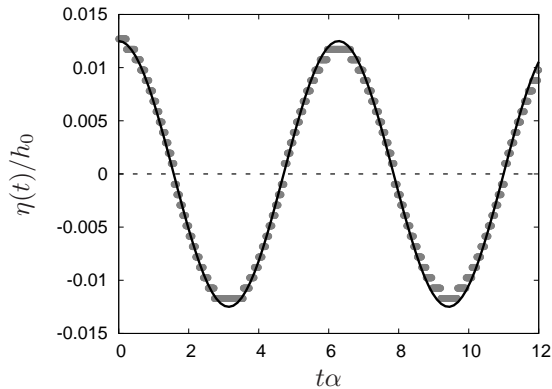}
\caption{\label{fig:stable} Time behavior of the perturbation 
maximum, $\eta(t)$, for $k=1$ and $h_0 / \lambda = 0.012$. 
The critical wave number  is
$k_c=0.9$. Numerical results (symbols) are compared with the 
prediction from linear theory (see Eq.~(\ref{eq:h_t})).
}
\end{figure}

The performance of the 
phase-field approach in the unstable regime predicted 
by linear theory both in the presence 
and in the absence of viscosity proved to be very good.
As discussed in Sec.~\ref{subsec:dispe}, for sufficiently 
large surface tensions and/or
sufficiently small differences between fluids density, 
a perturbation initially imposed
to the fluid interface may maintain 
its initial amplitude giving rise to the
dispersion relation (\ref{eq:om2}). 
The waves resulting from the balance between gravity and surface tension
are known as gravity-capillary waves. Our aim here is 
to verify their dispersion relation.\\
To do that, we have fixed the parameters to obtain a 
critical wavenumber of order one. 
For $\mathcal{A}g=0.008$ (model units) and the same $\sigma$ as 
in the unstable case, one has $k_c=0.9$. The
first accessible wavenumber is thus stable and should evolve 
in time according to (\ref{eq:om2}).
However, the geometrical/computational  configuration used 
in the unstable case did not produce
sufficiently accurate results. In particular, using the same 
domain aspect ratio $L_x/L_y = 1$ 
and the same ratio between
perturbation 
amplitude and perturbation wave-length we found a dynamics too dissipative with respect to what is 
expected.  In the absence of viscosity, 
dissipation arises in the phase field formulation due to the sole contribution proportional to $\gamma$
in Eq.~(\ref{eq:etot}). The latter parameter has been taken sufficiently small to ensure a negligible effects 
inside a period of oscillation. The specific value was $\gamma = 6.25 \times 10^{-5}$.
To avoid spurious dissipation, as that induced by nonlinear effects, we reduced the amplitude of the 
initial perturbation with respect to the unstable case. Also, we increased the size of the periodicity box
along the gravitational direction in a way to reduce possible spurious contribution arising
from the upper/lower part of the computational domain where instabilities, not present in the unstable case,
might now develop. 
The above choice on the amplitude of the inital perturbation
implies a consequent reduction 
of $\epsilon$. The following set 
of parameters have been used: $\epsilon=0.008$ ,  $L_x/L_y = 1/4$ and a resolution $Nx \times Ny$ 
of $256 \times 4096$. For an initial perturbation on $k=1$, its initial amplitude $h_0$ has been chosen 
to have $h_0/\lambda=0.012$ and $h_0/ \epsilon \sim 10$. 
The behavior of the maximum, $\eta(t)$, of the initial perturbation is shown as a function of time in  
Fig.~\ref{fig:stable}. 
The continuous line is relative to a sinusoidal with pulsation $\omega $ obtained from
(\ref{eq:om2}). The agreement between
theory and numerics is satisfactory both for the amplitude and for the pulsation.
Note the small reduction of $\eta(t)$, in one oscillation period: only 1 grid box over 4096.

\subsection{Weakly non-linear stage}\label{num_nl}
\begin{figure}
\centering
\includegraphics[scale=0.5]{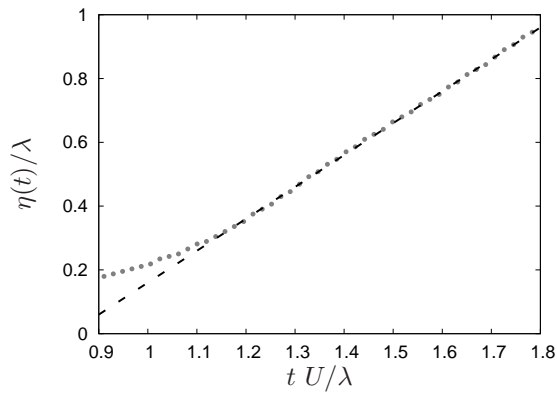}
\caption{\label{fig:h_nn} Time evolution of amplitude perturbation $\eta(t)$. 
The dots are our numerical results, the dashed line is the prediction
 by Eq.~(\ref{eq:U_nn}).}
\end{figure}

\begin{figure}
\centering
\includegraphics[width=2.5cm,height=5cm]{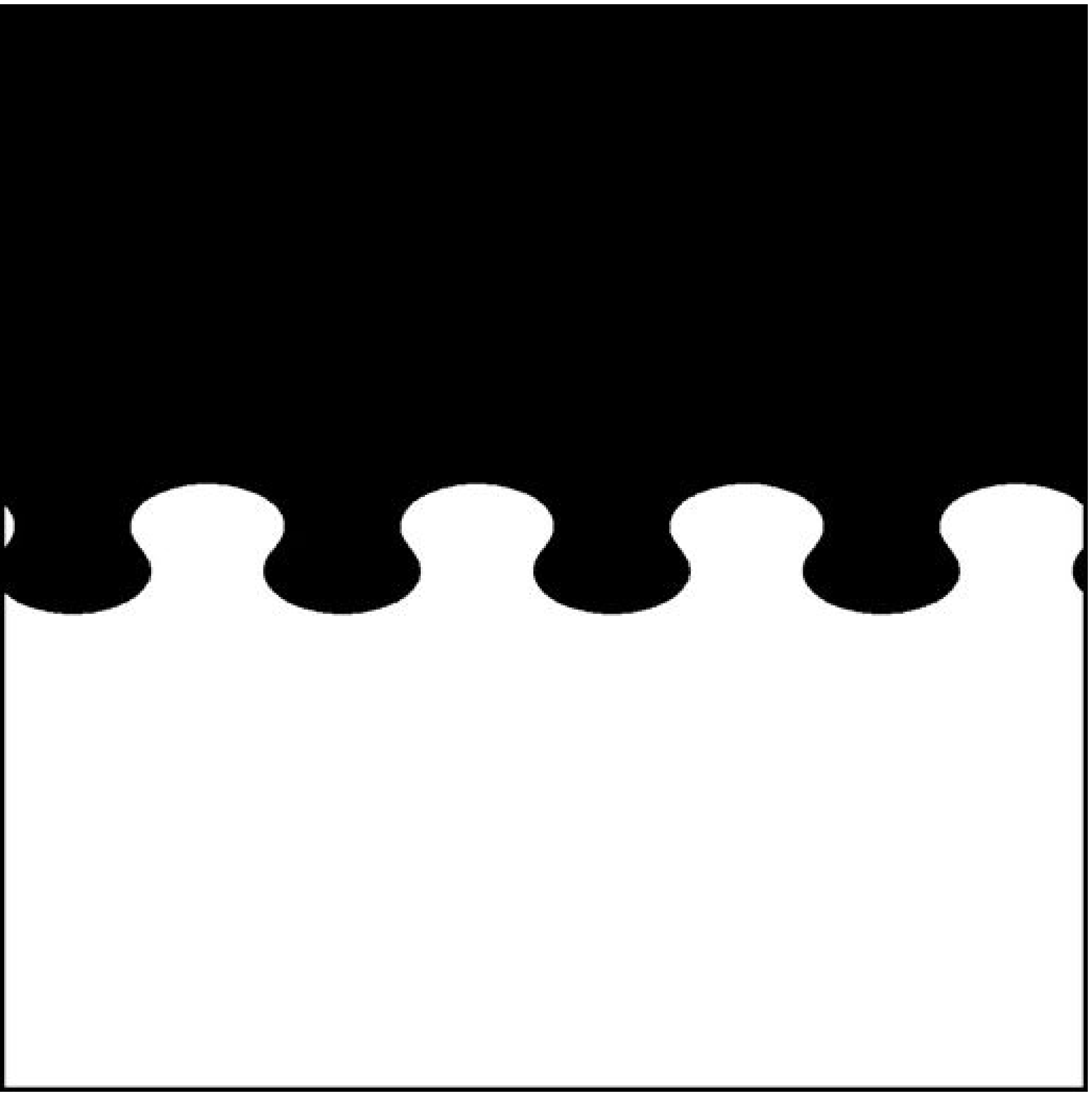}
\includegraphics[width=2.5cm,height=5cm]{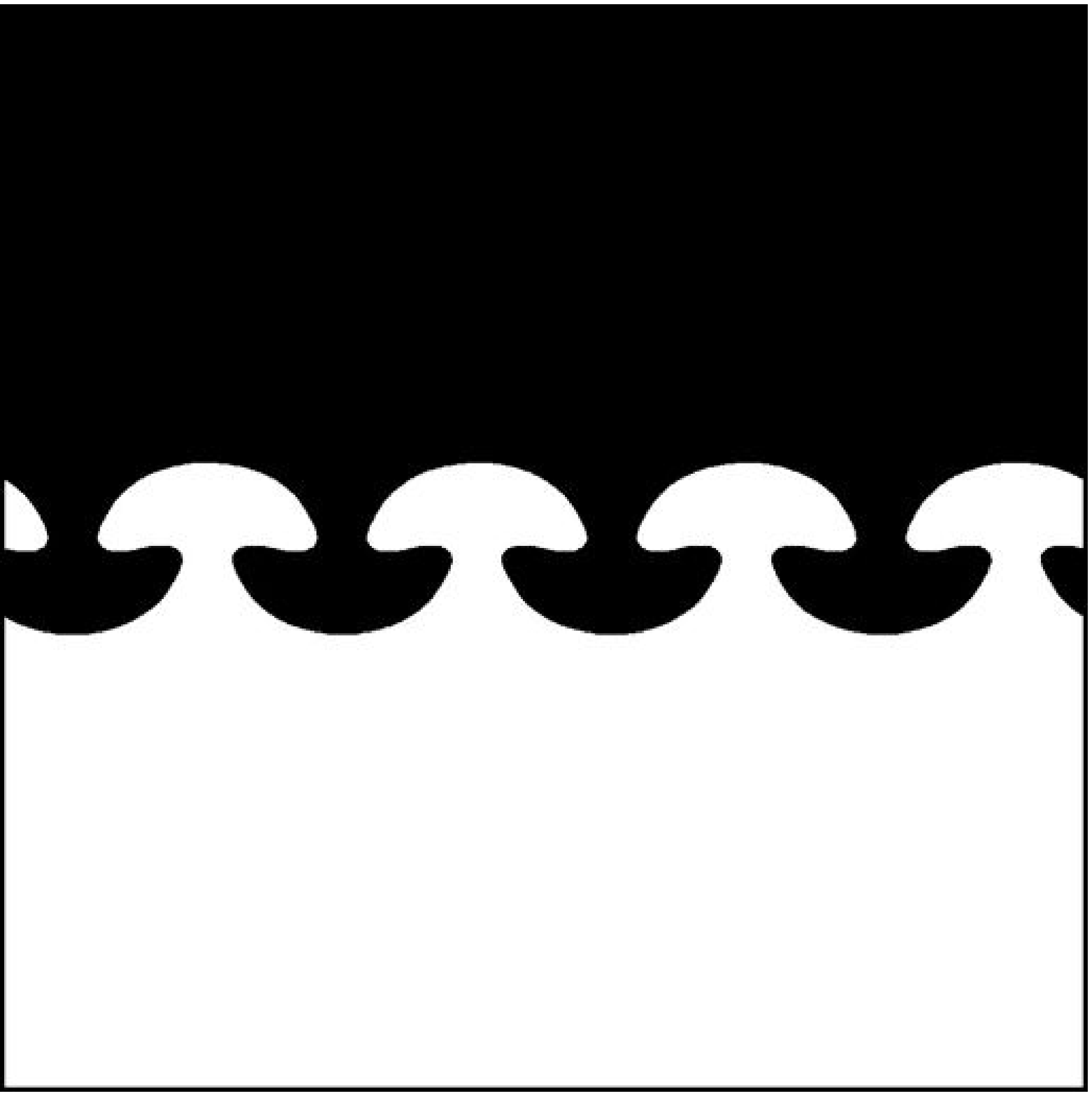}
\includegraphics[width=2.5cm,height=5cm]{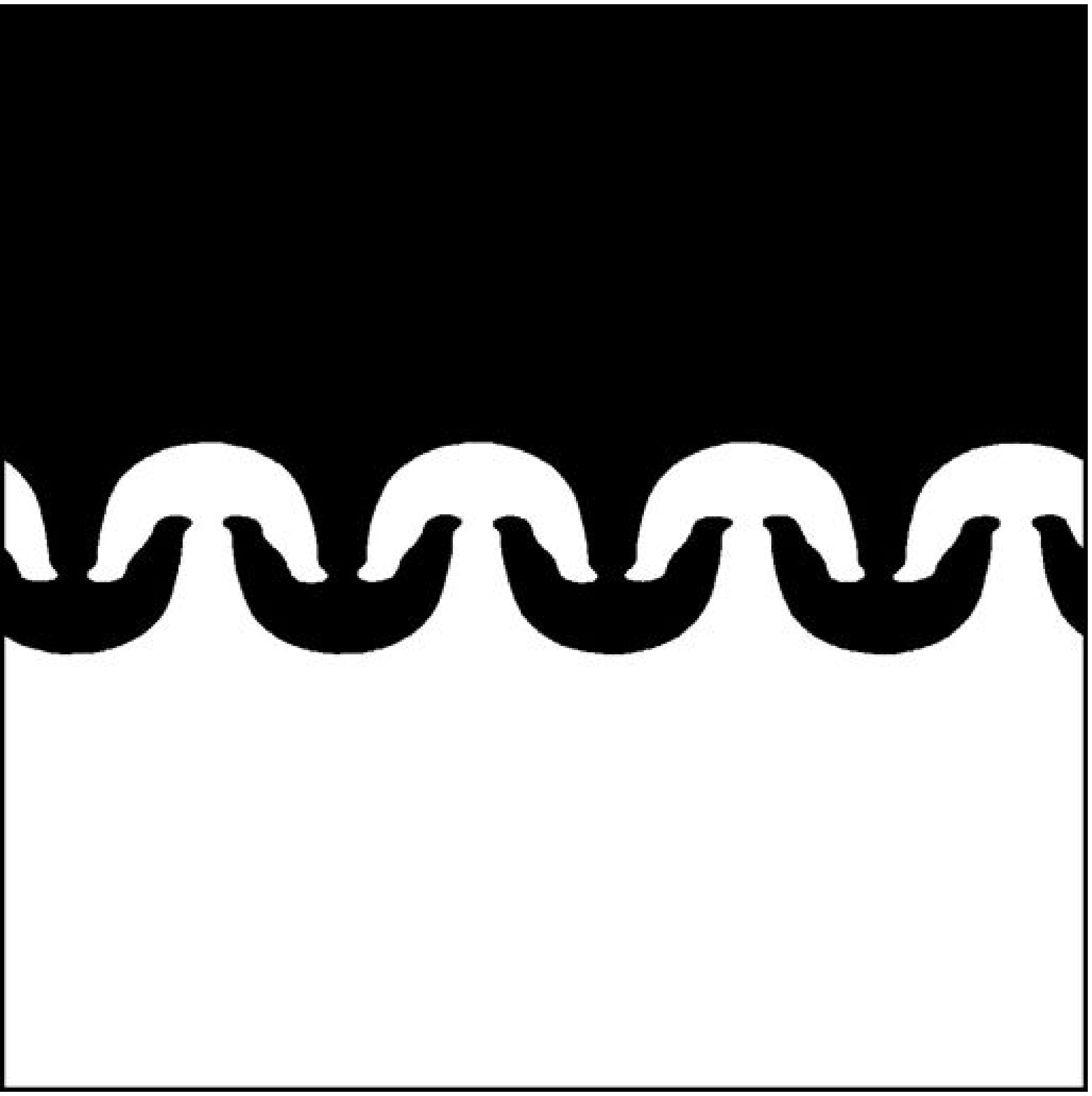}
\includegraphics[width=2.5cm,height=5cm]{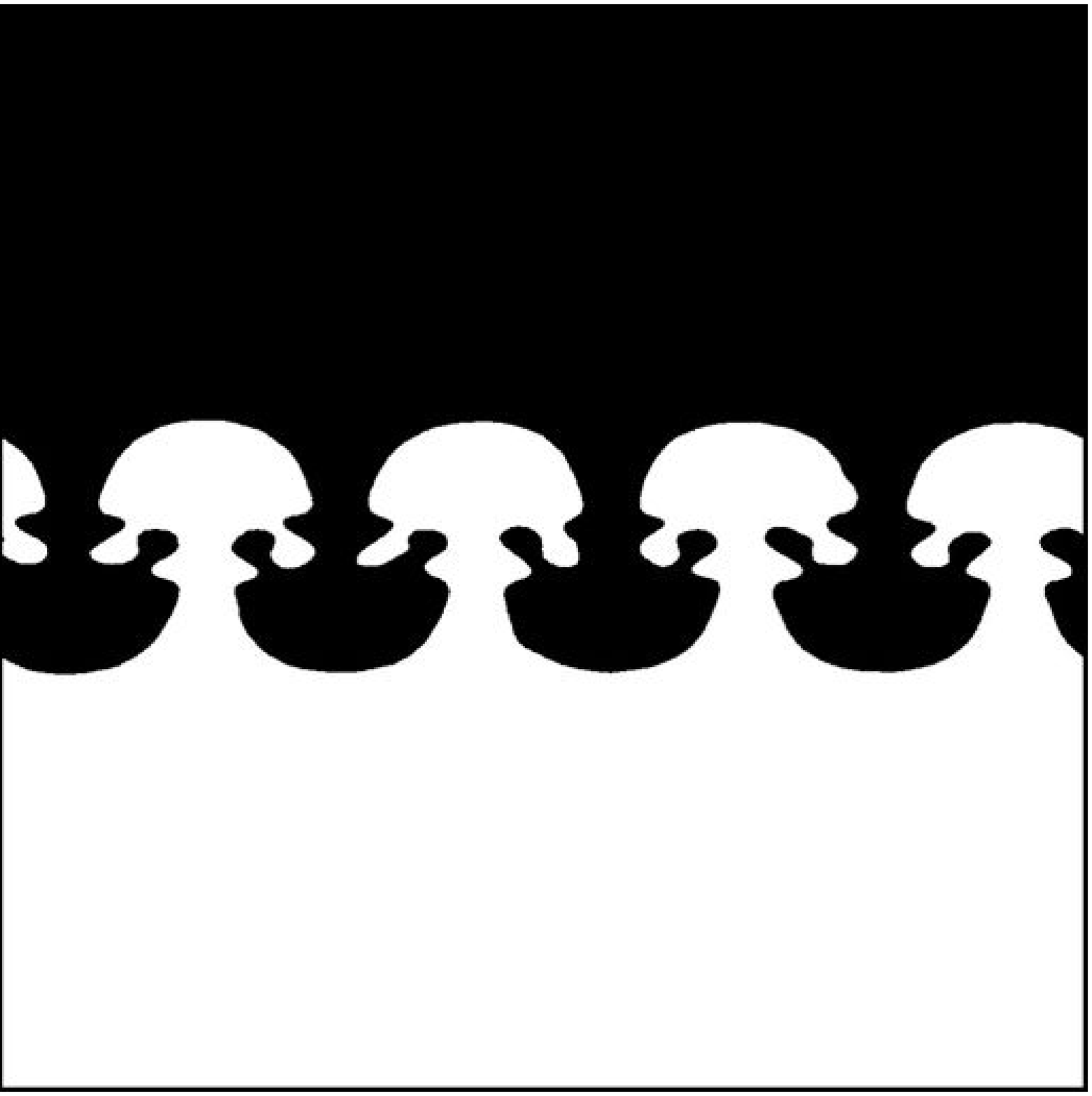}
\caption{\label{fig:plumes} 
Two-color snapshots of the  phase field. Black 
(white) corresponds to $\phi=-1$ ($\phi=1$). Frames are
equally spaced in time in the interval $1.2 < t U / \lambda < 1.8$ (see also Fig.~\ref{fig:h_nn}). }
\end{figure}

In this section we investigate the early stages of the nonlinear dynamics.
We focus on the rising/falling velocity of plumes
in the limit of small Atwood numbers when spikes and bubbles are known to coincide.
The theoretical prediction for the terminal velocity is reported in 
Appendix \ref{sec:nnl}. Our aim here is both to verify the existence of a regime
characterized by a costant `terminal' velocity and, secondly, to compare the prediction (\ref{eq:U_nn})
for such terminal velocity with our numerical data.\\
The physical parameters are chosen to magnify the effect of the surface
tension on the terminal velocity. This happens when 
the wavenumber $k$
of the initial pertubation (still supposed unimodal) is slightly below $k_c$.
Here we choose   $\mathcal{A}g$ and $\sigma$  such that  $k_c=4.004$ and thus 
look at the dynamics associated to the 
wavenumber $k=4$. The initial perturbation has an amplitude $h_0 / \lambda =0.06$;
the initial dynamics is thus linear.
Although we are interested to investigate the case of zero viscosity, in order to prevent
numerical instabilities we add a small viscosity 
$\nu=2\times 10^{-5}$ (model units). 
In  Fig.~\ref{fig:h_nn} the perturbation amplitude is shown as a function of 
time: symbols correspond to our numerical data and the dashed line is the prediction (\ref{eq:U_nn}).
A good agreement is found between numerics and theory in the range 
$1.2 < t U / \lambda < 1.8$. 
At larger times, neighboring plumes start to interact
and the arguments leading to (\ref{eq:U_nn}) do not apply any longer.
In Fig.~\ref{fig:plumes}
we show some snapshots of the evolution of the two fluids. Figures 
are equally spaced in time in the interval $1.2 < t U / \lambda < 1.8$.
Black corresponds to $\phi=-1$; white to $\phi=1$.
Their shape is similar to that experimentally observed.
Note the aforementioned spike/bubble symmetry
corresponding to the up-down symmetry of our original evolution equations.
by  \cite{WNJ01}.

\section{Conclusions and perspectives} \label{conclu}
In this paper we showed  that the phase--field model  provides  
a valuable numerical instrument for the study of immiscible, convective hydrodynamics. As a testground for this model, we have considered the Rayleigh--Taylor 
instability. Numerical results compare very well with known analytical results 
both  for the linearly stable and unstable case, 
and for the weakly nonlinear stages of the latter. \\
All these results are very encouraging in view of the 
next important step that is the
the numerical simulation of immiscible RT turbulence. There, the interplay 
of all the fundamental mechanisms that we have illustrated here 
(instabilities and wave propagation)
is expected to give rise 
to a small-scale emulsion-like phase dominated by gravity-capillary waves 
and by a large-scale
hydrodynamic range of scales where classical Kolmogorov 
turbulence should appear. This theoretical suggestion still awaits numerical 
confirmation, and the phase--field model provides the appropriate method
to pursue this goal. 

\begin{acknowledgements}
We acknowledge useful discussions with  Hekki Haario. 
AM and LV have been partially supported by PRIN 2005 project n.~2005027808 
and by CINFAI consortium (AM). LV acknowledges support from 
From Discrete to Continuous models for Multiphase Flows TEKES  project n. 40289/05. 
\end{acknowledgements}

\begin{acknowledgements}

\end{acknowledgements}

\appendix
\section{Bounds for the perturbation growth-rate in the presence of viscosity }\label{sec:l_vu}
The effect of viscosity is to 
reduce the perturbation growth-rate. 
However it does not remove the instabilities. 
Analytically,  
it is more difficult to consider the effect of viscosity with respect to
surface tension \cite[see Eq.~(115) at page 443 of][]{C61}. 
Nonetheless, 
it is possible to determine a lower and an upper bound to the 
growth-rate $\alpha_\nu$. These bounds are the solutions to the following 
equations  \cite[][]{MMSZ77}: 
\begin{eqnarray}
 \alpha_\nu^4 + 2\nu k^2 \alpha_\nu^3 
+ (\nu^2k^3-\frac{\alpha^2}{k})k\alpha_\nu^2 -(\nu^2 k^3 + \frac{\alpha^2}{k})\nu k^3\alpha_\nu  - (\nu^4 k^6 - \frac{\alpha^4}{k^2})k^3 &=& 0 \label{eq:low}\\
 \alpha_\nu^2 + 2\nu k^2 \alpha_\nu -\alpha^2 &=& 0  \quad.\label{eq:up}
\end{eqnarray}
where $\alpha$ is the growth-rate in the inviscid case (see Eq.~(\ref{eq:alpha})).  
The solution of Eq.~(\ref{eq:up}) is:
\begin{equation}
\alpha_\nu = -k^2\nu + \sqrt{ k^4 \nu^2 + \alpha^2}
\end{equation} 
while only a numerical solution is available for Eq.~(\ref{eq:low}).\\
The goodness of those upper and lower bounds are numerically investigated 
in Sec.~\ref{num_nu} by means of the phase-field model.
\section{Models for the terminal bubbles/spike velocities in the weekly nonlinear regime}\label{sec:nnl}
Substantial deviations from the linear theory are observed 
when the perturbation amplitude reaches a size of the 
 order of 0.1$\,\lambda$~-~0.4$\,\lambda$ \cite[][]{S84}. \\
In that case the perturbation evolution is nonlinear. 
 Then the disturbance grows non-linearly  
and the interface  starts to deform.  
Indeed, at least for finite values of $\mathcal{A}$, 
the interface can be divided into spikes corresponding to the regions where  
the heavier fluid penetrates into the lighter one, and 
 bubbles associated to those regions where 
lighter fluid rises in the heavier one. 
The roll-up of vortices produces a mushroom-type 
shape for bubbles and spikes \cite[see, for instance,][]{WNJ01}. 
When the fluid densities are similar  
(corresponding to our case $\mathcal{A} \ll 1$)  
spikes and  bubbles coincide and approach a constant and equal velocity. 
In both cases, the exponential growth of the velocity perturbation amplitude 
characterizing the linear phase of the evolution  is
replaced by a linear-in-time behavior \cite[][]{WNJ01}.  
Two models are available to describe this stage: 
the drag-buoyancy model \cite[][]{AHOS95} 
and the ``Layzer model'' \cite[][]{L55,G03,YH06}. 
The former model describes bubble and spike motion by balancing the 
buoyancy and drag forces and it assumes that this velocities reach a 
constant values for sufficiently long times. 
The latter model uses an expansion of the perturbation amplitudes and conservation 
equations near the tip of  bubbles 
and spikes. 
This 
approach has been first 
applied to the fluid-vacuum interface ($\mathcal{A}=1$) 
\cite[][]{L55} and then extended to arbitrary Atwood 
numbers \cite[][]{G03} and to include 
the surface tension contribution \cite[][]{YH06}. 
According to the latter study, 
in our case (bidimensional flow, immiscible fluids and small Atwood number)
one expects that the terminal bubble 
and spike velocity  be equal to \cite[][]{YH06}:
\begin{equation}\label{eq:U_nn}
U(t\rightarrow \infty)=  
\sqrt{\frac{2}{3}\mathcal{A}\frac{g}{k}- 
\frac{1}{9}\frac{\sigma}{\rho_2 + \rho_1}k}\quad.
\end{equation}
This expectation is numerically tested, in Sec.~\ref{num_nl}, by exploiting 
the phase-field method.

\end{document}